\documentclass[fleqn,usenatbib]{mnras}
\usepackage{newtxtext,newtxmath}
\usepackage[T1]{fontenc}
\DeclareRobustCommand{\VAN}[3]{#2}
\let\VANthebibliography\thebibliography
\def\thebibliography{\DeclareRobustCommand{\VAN}[3]{##3}\VANthebibliography}

\usepackage{graphicx}	
\usepackage{amsmath}	
\usepackage{caption}

\def\swift{{\it Swift}}
\def\xmm{{\it XMM-Newton}}

\def\ngc{{NGC~6814}}
\def\iras13{{IRAS~13224--3809}}
\def\1h07{{1H~0707--495}}
\def\izw1{{I~Zw~1}}

\def\deg{^{\circ}}

\def\cm{{\rm\thinspace cm}}
\def\erg{{\rm\thinspace erg}}

\def\ga{{\rm\thinspace gauss}}

\def\keV{{\rm\thinspace keV}}
\def\km{{\rm\thinspace km}}
\def\Rg{{\rm\thinspace R_g}}

\def\Msun{\hbox{$\rm\thinspace M_{\odot}$}}

\def\s{{\rm\thinspace s}}

\def\ps{{\rm\thinspace s^{-1}}}

\def\kmps{\hbox{$\km\ps\,$}}

\def\pscm{\hbox{$\cm^{-2}\,$}}

\title[An SED Variability Study of NGC 6814]{A Spectral Energy Distribution Variability Study of the Eclipsing AGN NGC 6814}

\author[L. K. Pothier-Bogoslowski et al.]{
L. K. Pothier-Bogoslowski,$^{1}$\thanks{E-mail: lucienne.pothier-bogoslowski@smu.ca (LKPB)}
L. C. Gallo,$^{1}$
A. G. Gonzalez, $^{1}$ M. Z. Buhariwalla, $^{1}$  J. M. Miller $^2$
\\
$^{1}$Department of Astronomy and Physics, Saint Mary's University, 923 Robie Street, Halifax, B3H 3C3, Canada\\
$^2$Department of Astronomy, University of Michigan, 1085 South University Avenue, Ann Arbor, MI 48109-1107, USA\\
}
\date{Accepted XXX. Received YYY; in original form ZZZ}

\pubyear{\the\year{}}

\begin{document}
\label{firstpage}
\pagerange{\pageref{firstpage}--\pageref{lastpage}}
\maketitle

\begin{abstract}
The local Seyfert 1.5 active galactic nucleus (AGN), NGC 6814, is known to exhibit complex variability, eclipses, and even changing-look behaviour.  In this work, we utilize optical-to-X-ray data obtained over 10-years with the Neil Gehrels \swift{} Observatory to examine the short-term (i.e. daily) and long-term (yearly) variations in the spectral energy distribution (SED). This includes \textcolor{black}{three epochs of high-cadence monitoring (2012, 2016, and 2022), as well as two single observations (2019 and 2021)}. Model-independent methods of examining the variability suggest that the three monitored epochs exhibit distinct behaviour.  X-ray weakness in 2016 can be attributed to the previously studied eclipses, while similar behaviour in 2012 is associated with continuum changes and slight neutral absorption.  The multi-epoch SED models are consistent with a black hole \color{black}{(${\log (M_{BH}/\Msun{})\approx 7.6}$)} \color{black} that is accreting between $\dot{m}\approx0.01-0.1$.  While the corona (primary X-ray source) is compact, all epochs are better fit with an accretion disk inner radius that is much larger than the innermost stable circular orbit, implying the possibility of a non-standard accretion disk or central structure in \ngc{}.
\end{abstract}




\section{Introduction}
Active galactic nuclei (AGN) are unique astrophysical objects as they emit nearly equal amounts of radiation in each energy band from radio to X-rays and $\gamma$-rays.  The primary AGN emission is generated in the central region as an accretion disk of material falls toward the supermassive black hole (SMBH). A standard disk structure that is considered to be geometrically thin and optically thick, emits as a multi-temperature black body, which peaks in the UV and cools with increasing distance ($R$) following a $T\propto R^{-3/4}$ relation \citep[e.g.][]{Shakura1973,Novikov1973}.  The characteristic X-rays are produced in the central corona surrounding the inner disk as UV photons from the accretion disk collide with the hot electrons in the corona and are inverse Compton scattered  \citep[][]{Haardt1993}.

With increasing distance from the central region, the disk cools and emits in the optical and infrared \citep[][]{Shakura1973,Novikov1973}.  Broad UV, optical, and IR emission lines are emitted from the broad-line region (BLR) \citep[e.g.][]{Kaspi2000} in the outer disk and narrow optical / IR emission lines from the narrow-line region (NLR) \citep[e.g.][]{Netzer1993} on galaxy scales.    The distant dusty torus is responsible for IR emission as it reprocesses and obscures the primary AGN emission \citep[e.g.][]{Nenkova2008}.  All the AGN emission can be modified by ionized material in the form of winds and obscurers \citep[e.g.][]{Gallo2023}.  Additionally, radio and $\gamma$-rays can be prevalent in the presence of jets \citep[e.g][]{Lister2020}.

All these various components are connected either gravitationally or radiatively.  However, most AGN regions are spatially unresolved and to probe the interplay between them we examine the spectral energy distribution (SED).  An SED analysis can provide insight into the origin and behaviour of AGN emission mechanisms like the soft excess and complex properties such as the bolometric correction and accretion rate \citep[e.g.][]{Grupe2010,Mehdipour2011,Done2012,Jin2012}.

The AGN components are also spatially separated by large distances.  Studying the SED variability allows us to observe the evolution of accretion disk properties and examine how they propagate to other components \citep[e.g.][]{Vasudevan2009,Mehdipour2011,Papadakis2016,Buisson2017,Tripathi2020}. Specific studies of X-ray weakness in AGN are important to demonstrate that the observed differences in intrinsic X-ray weakness and X-ray obscuring events can be subtle and require deeper examination through SED analysis \citep[e.g][]{Leighly2007,Zhang2023}.

\ngc{} \citep[$z=0.005223$][]{Springob2005} is classified as a Seyfert 1.5 \citep[][]{Veron2006}. From optical to X-ray, it is highly variable on both long and short time scales \citep[e.g.][]{Gonzalez2024}.  On more than one occasion, it has exhibited eclipsing behaviour \citep[][]{Leighly1994,Gallo2021,Pottie2023}, and there is evidence that it is a changing-look AGN, altering between absorbed and unabsorbed states \citep[][]{Sekiguchi1990, Jaffarian2020}. The high degree of variability, evidence of changing absorbers, brightness, and proximity make \ngc{} a good candidate to study the SED properties. 

In this work, we examine the long (over ten years) and short-term (daily) optical-to-X-ray SED variability of \ngc{}, examining changes in the continuum as well as absorption. In Section \ref{obs} we describe the observations and data processing. In Section \ref{lc} we examine the variability in a model-independent manner. SED fitting of the average yearly and daily spectra is carried out in Sections \ref{model} and  ~\ref{all_daily}, respectively.  In Section \ref{cor} we describe the correlations and variability seen in SED model parameters.  The results are discussed in Section \ref{disc} and conclusions are presented in Section \ref{conclusion}.

\section{Observations and data reduction}\label{obs}
This work was completed using data from The Neil Gehrels \swift{} Observatory \citep[][]{Gehrels2004} from both the X-ray Telescope (XRT) \citep[][]{Burrows2005} in Photon Counting (PC) mode and the UltraViolet/Optical
Telescope (UVOT) \citep[][]{Roming2005}. All observations of \ngc{} were used, including high-cadence monitoring in 2012, 2016, and 2022, as well as short, single observations in 2019 and 2021. Table \ref{tab:obs} shows a complete list of observations and the X-ray and UVW1 light curves are shown in Figure \ref{fig:lc_all}. Light curves for 2019 and 2021 are not shown since those observations lasted only 1-2 days. We note that there are an additional 4 observations in 2016 (00081852001-00081852004) that were ignored because the XRT was operated in Windowed Timing (WT) mode. 

The XRT spectral  and light curve data were processed using the \swift{-XRT} Product Builder \footnote{\url{https://www.swift.ac.uk/user_objects/}} \citep[][]{Evans2009} in the 0.3-10 keV band, and binned by observation. In total there were 76, 60, and 67 days of observations for 2012, 2016, and 2022, respectively (1 observation in 2019 and 2 in 2021). The spectra were combined into yearly averages for the 2012, 2016, and 2022 observations, using the \textsc{addascaspec} task from \textsc{ftools} in \textsc{heasoft} version 6.32, in addition to roughly "daily" spectra created for 2016 and 2022. Since 2012 was in a dimmer state, the data were combined in larger temporal bins. The dashed vertical lines in the top panel of Figure \ref{fig:lc_all} indicate the different segments created for 2012, these are made such that the first two segments during the high flux intervals (H1-H2) contain 4 and 3 days of observations, the low flux intervals (L1-L4) each contain 10 days of observations, and the intermediate intervals (I1-I3) contain between 9 and 10 days of observations. 

The UVOT observations were all taken in imaging mode with the exception of the UVW1 filter for observation IDs 00032477223 and 00032477224 which were done in event mode. The data were processed as described in the \swift{-UVOT} Analysis Guide \footnote{\url{https://www.swift.ac.uk/analysis/uvot/mag.php}} using the \textsc{uvotsource} task from \textsc{ftools} with a source region of 5 arcseconds and background region of 15 arcseconds, with visually determined positions. The UVOT fluxes are averaged according to the same groupings described for the X-ray data, and then converted into spectral files using the \textsc{ftools} task \textsc{ftflx2xsp}.

\begin{table*}
    \centering
    \caption{The \swift{} observations of \ngc{} used in this work. Column (1) lists the epoch year, and Columns (2) and (3) show the start and end dates of the monitoring, respectively. Column (4) shows the observation IDs included in the epoch, and Column (5) shows the UVOT filters used in the observations of each epoch.}
    \begin{tabular}{c|c|c|c|c|c}
    \hline
    (1)&(2)&(3)&(4)&(5)&(6)\\
        Epoch & Start Date & End Date & Observation IDs & Days Observed & UVOT Filters \\ \hline
        2012 & 2012-06-02 & 2012-09-06 & 00032477002-00032477085 & 76 & V, UVW1 \\
        2016 & 2016-03-29 & 2016-07-08 & 00032477086-00032477221& 60&U, UVW1, UVW2, UVM2 \\
        2019 & 2019-05-09 & - & 00342477222 & 1& U \\ 
        2021 & 2021-09-29 & 2021-10-05 & 00032477223-00032477224 &2 & B, U, V, UVW1, UVW2, UVM2 \\
        2022 & 2022-08-28 & 2022-11-10 &00032477224-00032477255 &67& B, U, V, UVW1, UVW2, UVM2 \\ & & & + 00015314001-00015314240 & & \\ \hline
    \end{tabular}
    \label{tab:obs}
\end{table*}

\section{Model-independent examination of the UV-to-X-ray variability}\label{lc}

Prior to fitting the UV-X-ray spectral energy distribution in Section \ref{model}, we examine the variability in the various energy bands in a model-independent manner.

\subsection{The X-ray and UV light curves}
\begin{figure*}
	\includegraphics[width=\textwidth]{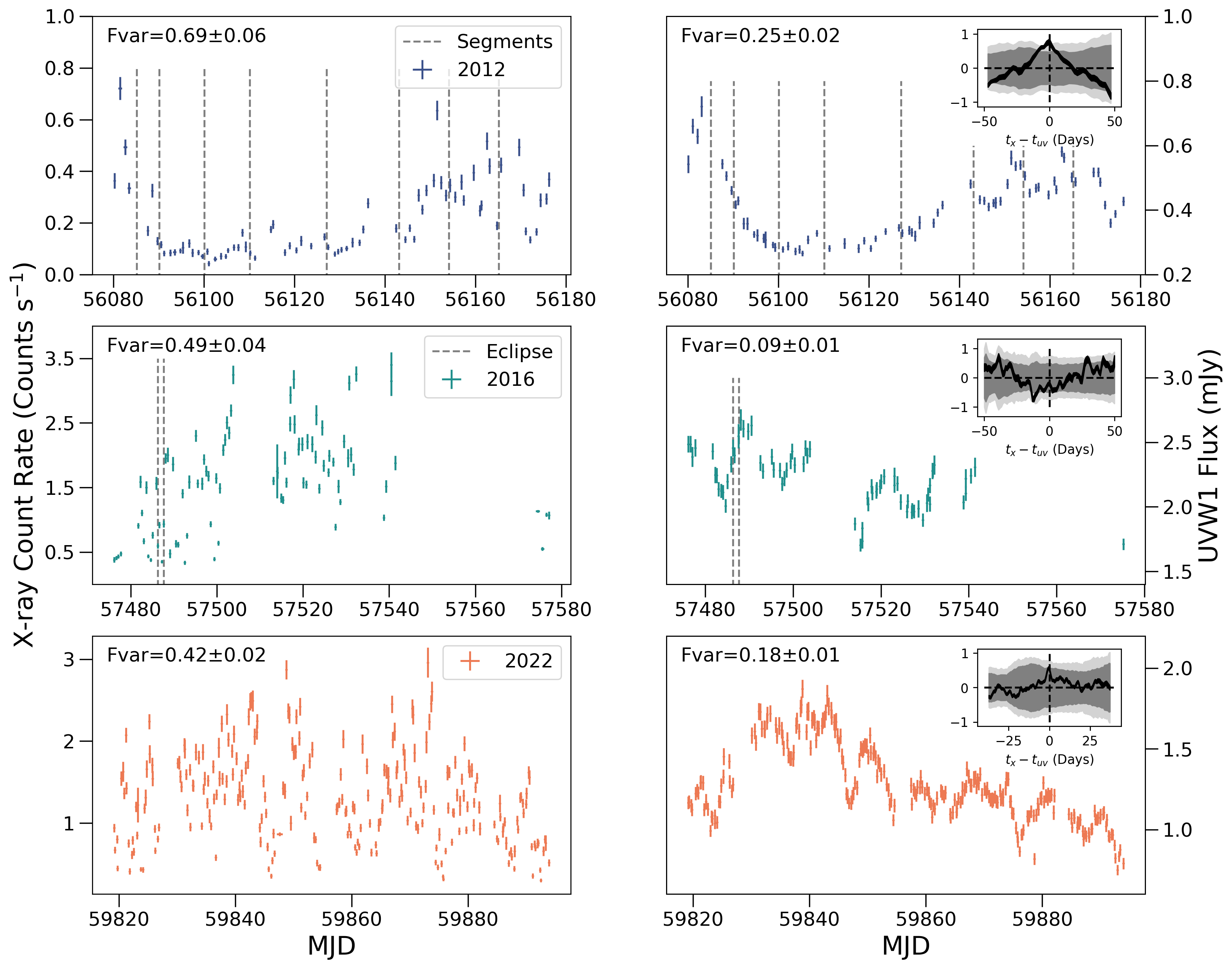}
    \caption{Left: \swift{} XRT light curves binned by observation for each of the three monitored epochs, 2012, 2016, and 2022. Grey dotted lines deliniate the segments in 2012 and the eclipse \citep[][]{Pottie2023} in 2016. Right: \swift{} UVOT light curves from the UVW1 filter, binned by observation for each of the indicated epochs. Insets: The discrete correlation function for each pair of light curves with 90 and 99 percent significance contours plotted in dark and light grey. The fractional variability is given in the top left corner of each plot.}
    \label{fig:lc_all}
\end{figure*}

The X-ray and UVW1 light curves for 2012, 2016, and 2022 (Figure \ref{fig:lc_all}) depict a high degree of variability on hourly-daily time scales across the duration of the 60-76 day monitoring campaigns. Based on the count rates, 2012 represents a relatively low-flux state; 2016 represents a relatively bright state; and 2022 is \textcolor{black}{considered to be in a ``typical’’ state}.

The discrete correlation functions \citep[][]{Edelson1988} \textcolor{black}{calculated between the X-ray and UVW1 light curves are shown in the inset panels of Figure \ref{fig:lc_all}}. We also present the 90 and 95 percent significance contours on the DCF plots.  The X-ray and UV light curves are well correlated with near zero delay between the signals.  This is consistent with the thorough correlation investigations of the 2012 and 2022 light curves by \citet{Troyer2016} and \citet{Gonzalez2024}, which measure lags of less than a day.  Interestingly, the 2016 light curves do not appear to be well correlated. This is not surprising when we recall that the X-rays were eclipsed \citep[][]{Gallo2021}, perhaps frequently \citep[][]{Pottie2023}, during this epoch. Indeed, if we neglect the data \textcolor{black}{effected by the confirmed and possible eclipses, $\le 57510$ days}, the DCF does show a peak at zero lag.  Given the short lag and good correlation between the light curves we do not consider time-delay effects \citep[e.g.][]{Kammoun2024,Kammoun2021} when fitting the SED in Section \ref{model}. 

We quantify the flux variability at each epoch by measuring the fractional variability  and its uncertainty \color{black} {as given by  \citet{Edelson2002}.} \color{black}The fractional variability for each light curve is listed in Figure \ref{fig:lc_all}. For stationary processes, one would expect that the fractional variability remains relatively constant over time \citep[e.g.][]{Vaughan2003, Alston2019}. This is not clearly the case for \ngc{}.  Considering the X-ray light curves, $F_{var}$ is similar in 2016 and 2022, but in 2012, $F_{var}$ is much higher.  Likewise, for the UV light curves the 2012 and 2022 are similar, while the variability is much lower in 2016.  While much longer and better sampled light curves are necessary to be more definitive, these differences in $F_{var}$ could be indicative of physical changes in the underlying processes creating the light curves (i.e. the SED).

\subsection{The UV-to-X-ray spectral shape}
\begin{figure}
    \centering
\includegraphics[width=\columnwidth]{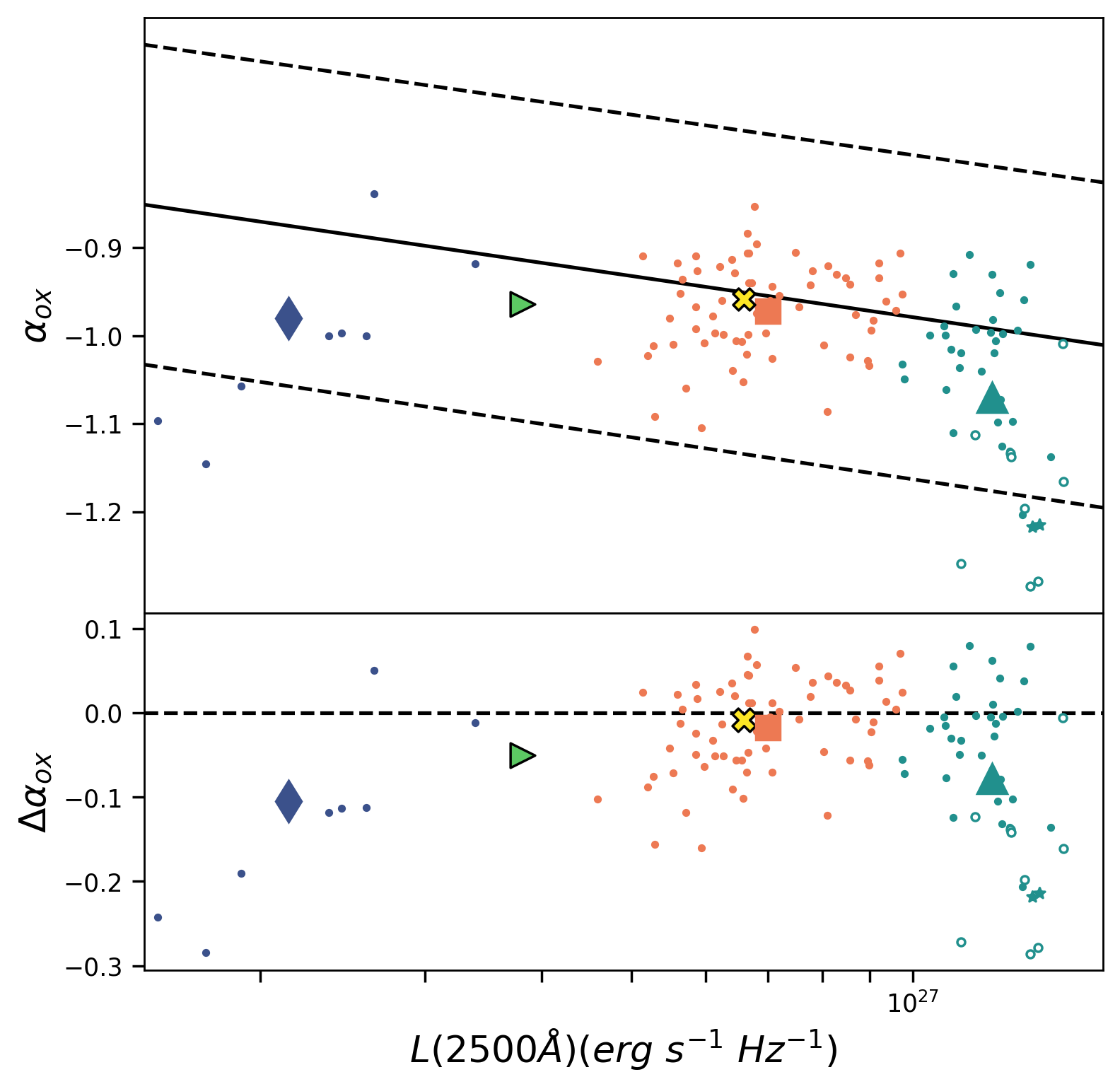}
    \caption{Upper panel: The $\alpha_{ox}$ measurements for the 2012 average and segmented spectra (blue diamond and points), the 2016 average and daily spectra (teal upward triangle and points), and the 2022 average and daily spectra (orange square and points). The 2016 eclipse and possible eclipse points are indicated by teal stars and open circles, respectively. The outlined shapes for 2018 (green right-pointing triangle) and 2021 (yellow cross) are calculated indirectly using average count rates. \textcolor{black}{The solid and dashed lines represent the expected relation from \citet{Nanni2017} and its upper and lower limits, respectively.} Lower panel: The X-ray weakness ($\Delta\alpha_{ox}$) measurements calculated from the expected relation.}
    \label{fig:aox}
\end{figure}

We examine changes in the SED shape by calculating the \color{black}{UV-to-X-ray spectral slope between 2500 \AA{} and $2~\keV$ \citep[$\alpha_{ox}$; ][]{Tananbaum1997} for each observation.}\color{black} This is calculated for each daily observation in 2016 and 2022, and for the segments in 2012.  This is also calculated for the annual average in 2012, 2016, 2019, 2021, and 2022 (Figure \ref{fig:aox}).

In instances when the 2500 \AA{} flux could not be directly measured from a UVW1 observation, the next closest adjacent filter was used and a UV spectral slope between that filter and 2500 \AA{} was assumed as outlined in \citet{Gallo2006}.  The UV spectral slope assumed here was $\alpha_u=-2$, based on the slope of the average UVOT spectrum measured in 2022, which included all filters.

In addition, due to the limited data available for 2019 and 2021, the counts are not sufficient to model the spectrum and measure the $2~\keV$ flux density ($f_{2\ keV}$) \ directly. Instead, a conversion factor between count rate and flux density is used for 2019 and 2021, based on the same conversion factor in 2022. These indirect calculations of $\alpha_{ox}$ are indicated by outlined shapes in Figure \ref{fig:aox}.

The measurements of $\alpha_{ox}$ for \ngc{} are compared to the $\alpha_{ox}-L_{2500\text{ \AA}}$ relationship from \color{black}\citet{Nanni2017}:
\begin{equation}\label{aox}
    \alpha_{ox}=(-0.155\pm0.003)\log(L_{2500\text{ \AA}})+(3.206\pm0.103)
\end{equation}

\color{black}
\noindent In doing so, we can examine the relative strength of X-ray to UV emission at each epoch by measuring $\Delta \alpha_{ox}$, which is the difference between the measured $\alpha_{ox}$ of \ngc{} and the expected value from \textcolor{black}{\citet{Nanni2017}} based on its luminosity (Equation \ref{aox}).  This measure of X-ray weakness is plotted in the lower panel of Figure \ref{fig:aox}.

In general, our observations seem to follow the trend found in \textcolor{black}{\citet{Nanni2017}} relatively well, with some notable excursions into X-ray weakness ($\Delta \alpha_{ox}<0$). In 2016, many of these X-ray weak epochs are associated with the eclipse \citep[][]{Gallo2021} or possible eclipses as identified by \citet[][]{Pottie2023} and defined as observations with a hardness ratio $>-0.2$ and \swift{} count rate $< 1.0$.  These possible eclipses are marked by open circles, with the confirmed eclipse days marked by stars. We note that the 2012 low state exhibits similar $\Delta \alpha _{ox}$ values to the 2016 eclipses.  This is intriguing as it may be indicating obscuration as a possible cause for the 2012 low state and motivates testing a partial coverer on the 2012 spectra, as described in Sections \ref{ave_2012} and \ref{segments_2012}.

\subsection{Spectral Principal Component Analysis}\label{pca}
To characterize the spectral variability in a model-independent way, we perform a principal component analysis (PCA) on each of the three monitored epochs. A PCA deconstructs a data set (matrix) into a set of eigenvectors called principal components, thereby reducing the dimensionality. The original data set can be reconstructed through linear combinations of the calculated eigenvectors. Each eigenvector then accounts for a percentage of the overall variance in the data, proportional to the associated eigenvalue. The variance from each principal component decreases with increasing eigenvalue number; the first principal component is the most significant \citep[][]{Feigelson2015}. For a spectral PCA, the initial data set is a set of time-resolved spectra \citep[e.g.][]{Parker2014,Buhariwalla2024}.
In this work we utilize the PCA code distributed by \citet{Parker2014}, which was ported from \textsc{python2} to \textsc{python3} by \citet{Buhariwalla2024}. 

\begin{figure*}
    \centering
\includegraphics[width=\textwidth]{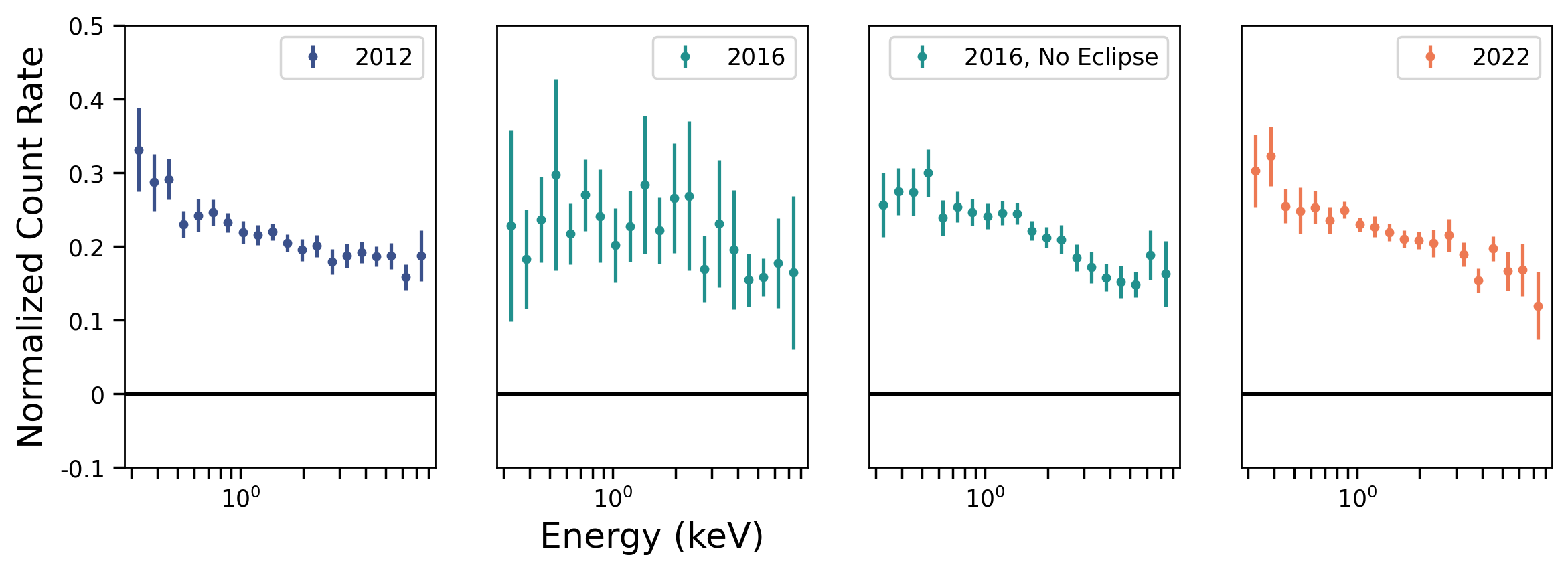}
    \caption{The first principal component (PC1) for each monitored epoch as labelled. In 2016, PC1 is calculated including (centre left) and excluding (centre right) data during the eclipses.}
    \label{fig:spec_pca}
\end{figure*}

We perform a PCA for each of the three monitored epochs, using the daily spectra as the input data set for 2016 and 2021, and the segmented spectra for 2012 (see Figure \ref{fig:lc_all}). We find that for each epoch only the first principal component (PC1) is significant, and it accounts for $\sim 93.7$, 43.1, and 57.4 percent of the total variance in 2012, 2016, and 2022, respectively. \textcolor{black}{A significant component is considered to be one whose eigenvalue clearly resides above the linear scatter.} PC1 for each epoch is displayed in Figure \ref{fig:spec_pca}. 

The shape of the PCA is comparable in 2012 and 2022, exhibiting diminishing variability toward higher energies.  The shape of PC1 in 2016 differs and appears rather flat.   Excluding the days associated with the eclipse \citep[][]{Pottie2023} and recomputing the PCA in 2016, we find that PC1 (now accounting for 68.1 percent of the variability) also resembles 2012 and 2022 (Figure \ref{fig:spec_pca}).  The difference originally noted could be from the presence of an eclipsing cloud in 2016.  Disregarding the eclipsing event, the underlying variability appears very similar at all three epochs.

The behaviour of decreasing variability with increasing energy is frequently seen in the X-ray PCA of AGN.  \citet{Parker2014} note that a power law  varying in normalization in the presence of a constant hard component could produce this shape.  \citet{Gallant2018} demonstrated that this PC shape can be reproduced when the power law varies in normalization and slope in a correlated manner.  This is consistent with the often observed steeper-when-brighter phenomenon seen in AGN X-ray spectra.

Overall, this initial examination of the SED variability highlights several interesting points.  Based on the PCA, at all three epochs, the primary variability is likely due to a power law varying in brightness and shape in a correlated manner, but with the presence of an obscuring medium in, at least, 2016.  In 2012, \ngc{} is in a low-flux state and often exhibits periods of X-ray weakness like in 2016.  However, these X-ray weak states might not be due to eclipses in 2012.  Finally, the differences in the fractional variability in 2012 compared to 2016 and 2022 point to physical changes in the driving mechanisms.

\section{Fitting the optical-to-X-ray SED at each epoch}\label{model}

\begin{figure*}
    \centering
    \includegraphics[width=\linewidth]{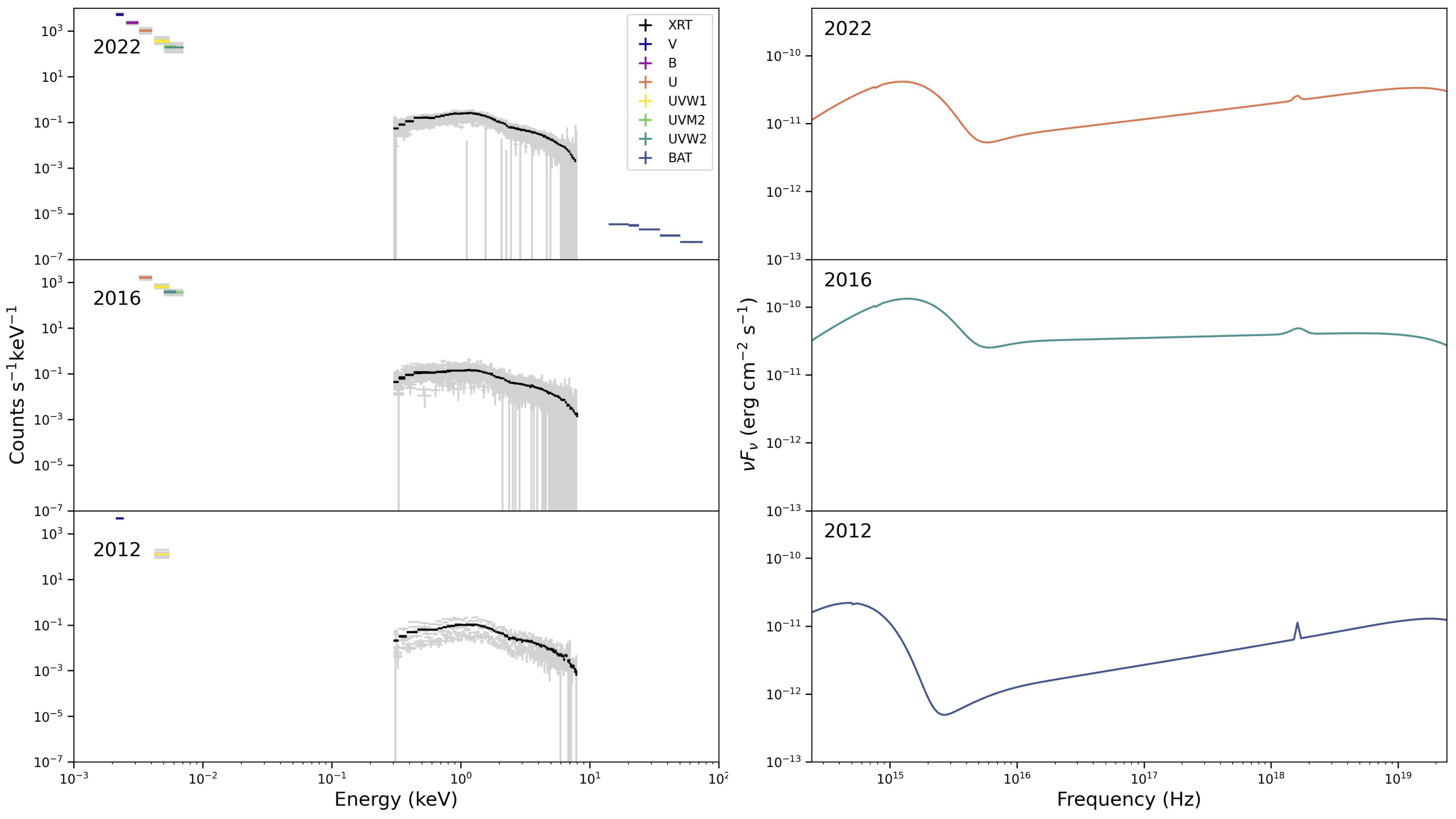}
    \caption{Left: The average SEDs and best fit model. The black and colourful points are the average data and best fit model, the grey are the daily/segmented spectra with the average model applied but not fit. \textcolor{black}{Right: The intrinsic $\nu F_\nu$ spectrum for each epoch average, all components not directly emitted by the central engine are removed.} }
    \label{fig:all_spec}
\end{figure*}

\begin{figure*}
    \centering
    \includegraphics[width=\textwidth]{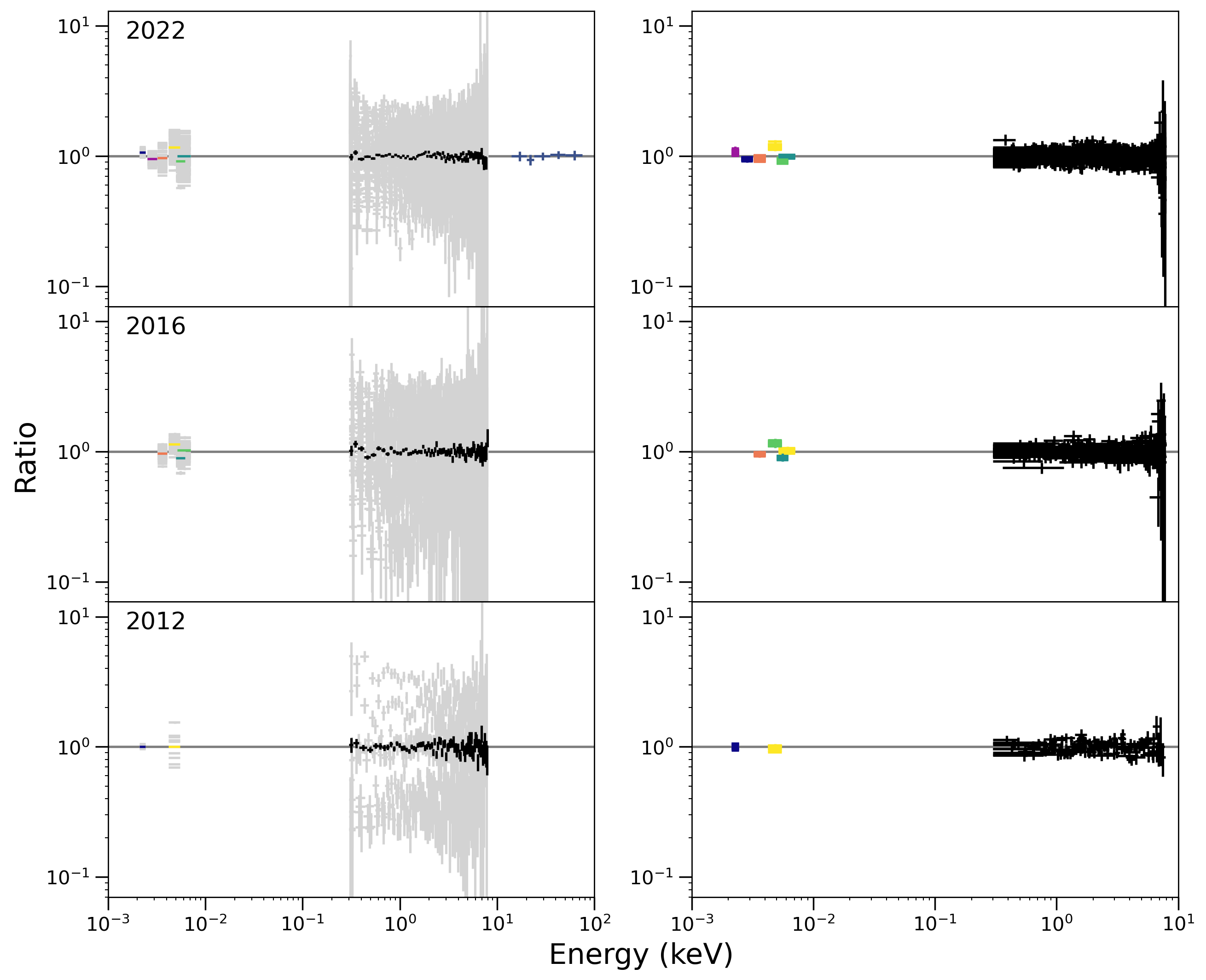}
    \caption{Left: The ratios from fitting the average SEDs (black and colourful points). The ratios from the daily/segmented spectra compared to the average model are overplotted (grey) highlighting the variability. Right: The ratios from fitting the daily and segmented SEDs. The X-ray data are binned up for viewing purposes.}
    \label{fig:ratios}
\end{figure*}

\textcolor{black}{In this section,} a self-consistent model is applied to the 2012, 2016, and 2022 spectral data to examine the variability in the SED.  Changes in the mean spectra are modelled in Sections \ref{ave_2022}, \ref{ave_2016}, and \ref{ave_2012}, and changes in the daily spectra are modelled in Sections \ref{daily_2022}, \ref{daily_2016}, and \ref{segments_2012}.  All the spectral data from the UVOT and XRT at each epoch are presented in Figure \ref{fig:all_spec} with ratios in Figure \ref{fig:ratios}.

\subsection{The broadband model}
The primary AGN continuum used to model the broadband SEDs is based on the combination of a standard accretion disk \citep[e.g.][]{Shakura1973} and thermal Comptonization \citep[e.g.][]{Sunyaev1979}.  No jet component is considered since \ngc{} is radio-quiet \citep[][]{Xu1999}.

The accretion disk is modelled as a disk black body assuming the general relativistic temperature profile of \citet{Novikov1973} and \citet{PageThorne1974} using a custom \textsc{xspec} \textcolor{black}{\citep{Arnaud1996}} model called \textsc{ntdisk} (Gonzalez et al. in prep). \textsc{ntdisk} includes the following set of input parameters: black hole mass, accretion rate, spin, inner and outer disc radii, inclination, colour-temperature correction factor, and comoving distance to the source. \color{black} The comoving distance to \ngc{} is fixed at 21.65 Mpc \citep[][]{Bentz2019}. Initially, black hole spin and inclination were free to vary, but did not improve the fits.  Therefore, the black hole spin was fixed to maximal and the inclination to $60\deg$, which is consistent with \citet{Gallo2021}. The colour–temperature correction factor is set to $f_{col}=1.7$, consistent with that of a black hole X-ray binary, \textcolor{black}{as was found by \citet{Gonzalez2024} to provide the best fit}. 

For the Comptonization component {\sc nthcomp} \citep[]{Zdziarski1996,Zycki1999} is adopted.  The seed electron temperature and black body temperature are fixed to 60 keV and 0.01  keV, respectively.

\textcolor{black}{The X-ray band in \ngc{} does seem to be occasionally modified} by a warm absorber and possibly a photoionized emitter \citep[e.g.][]{Gallo2021, Kang2023}. We model the warm absorber using the \textsc{spex} \citep{SPEX} model \textsc{xabs} \citep{Steenbrugge+2003} converted into an \textsc{xspec} readable table by \cite{Parker+2019}.   The turbulent velocity and the covering fraction were fixed to  $100 \km \s^{-1}$ and 1, respectively.

For the photoionized emitter, an \textsc{xstar} table \citep[][]{Kallman2001} was created.
In this {\sc xstar} table the ionizing continuum used was a simple power law with $\Gamma=2.2$. The turbulent velocity was fixed at 300\kmps, the covering fraction was fixed at 1, and the density was fixed at $n=10^{10}{\rm cm}^{-3}$. Two parameters, column density and the ionization parameter, were interpolated. The column density was logarithmically sampled at ten steps between $10^{20}$ and $10^{24}\cm ^{-2}$. Due to the quality of the \swift\ data, the column density was fixed to ${\rm n{H}}=1\times10^{24}\cm ^{-2}$. The ionization parameter is defined as $\xi= L/nr^2$ where $L$ is the ionizing luminosity between 1 and 1000 Ry, $n$ is the density of the emitting material, and $r$ is the distance between the emitting material and the ionizing source. The log of the ionization was linearly sampled ten times between 0 and 5. The table was calculated with 9999 energy bins from $0.1-50$\keV. The ionization and normalization were free to vary. 
In addition to the {\sc xstar} table, a Gaussian profile was included to account for the Fe K$\alpha$ emission line at $\sim 6.4$ keV.

In the UVOT band, a Balmer continuum \citep[][]{Grandi1982} is included to better fit the UV emission. The Balmer continuum temperature is frozen to 8000 K, we also apply \textsc{gsmooth} to the Balmer continuum with $\sigma_{6keV}=1\times10^{-4}\keV$ to account for broadening. \textcolor{black}{With an index of $\alpha=0$, and accounting for the position of the Balmer continuum at 3646 \AA, this is consistent with a} turbulent velocity of $\sim 10^4 \km \s^{-1}$ \citep[][]{Mehdipour2015}. In addition, we also account for contamination in the UVW2 and UVW1 band-passes due to C \textsc{iii]} at $\lambda=1908$ \AA{} and Mg \textsc{ii} at $\lambda=2798$ \AA{} \citep[see Fig. 1 of][]{Park+2017} by fitting narrow wavelength ranges of the Hubble Space Telescope (HST) Space Telescope Imaging Spectrograph (STIS) spectrum. The simple model consists of a power law continuum as well as one narrow and one broad Gaussian line profile to fit the emission line of interest. Best-fit model parameters are determined through a Markov Chain Monte Carlo (MCMC) run performed with the \textsc{emcee} Python package \citep{EMCEE} with 80 walkers for 11\,000 iterations (burning the first 1\,000) by maximizing a Gaussian likelihood function.

We also account for Milky Way extinction and reddening with \textsc{redden} and \textsc{phabs} \citep[values calculated by][]{Willingale2013}\footnote{\url{https://www.swift.ac.uk/analysis/nhtot/index.php}}. \textcolor{black}{Finally, we include a galaxy template for Sc type galaxies \citep[][]{Polletta2007} as was found by \citet{Gonzalez2024} to provide the best fit to the host galaxy emission}. 

The complete base model entered into \textsc{xspec} is: 
\textsc{redden * phabs * (xabs * (gsmooth * Balmer + zgauss + nthcomp + zashift * ntdisk) + Sc + xstar) + agauss + agauss + agauss + agauss}.

In addition to this base model, we test the fit improvement with extra absorption on the power law using both a partial coverer \color{black} {(an additional \textsc{xabs} component with variable covering fraction and ionization) }\color{black} and cold absorber \citep[\textsc{ztbabs},][]{Wilms2000} on each spectrum. The best-fit model parameters found for the simultaneous fit of the three monitored epochs are given in Table \ref{tab:model}.

Throughout this work we quantify the quality of our spectral fits using two statistical techniques. The X-ray fit statistic must be calculated using C-statistics \citep[][]{Cash1979}, which assumes a Poisson distribution of photon counts. This statistic does not suit the UVOT data which are better represented by a Gaussian distribution, thus, the $\chi ^2$ method is used for the UVOT data. To accurately compare the quality of the fit of multiple models on the same data set, we apply the corrected Akaike information criterion ($AIC_c$) \color{black} {as given by equation 1 in \citet{Akaike1974} and equation 4 in \citet{Hurvich1989}.}\color{black} 
To qualify a fit as "significantly" better than another we require a 95 percent confidence level which is equivalent to $\Delta AIC_c=6$ \citep[][]{Tan2012}.
We also note that the $AIC_c$ value is not the only factor by which we choose the best model, we often prioritize physicality of the fit parameters.

\subsection{The average 2022 spectrum}\label{ave_2022}
The model described above is first applied to the average 2022 spectrum.  This epoch possesses the most continuous sampling and is the only epoch to be observed in all of the UVOT filters.  In addition to the \swift{-XRT} and UVOT spectra, we utilize the \swift{} Burst Alert Telescope (BAT) \citep[][]{Barthelmy2005} spectrum from the 105-month hard X-ray survey  \citep[][]{Oh2018} to confirm our model is consistent at $E > 10\keV$.  To account for cross-calibration between BAT and XRT, we apply an additional free multiplicative constant to the BAT data which reaches a value of $\sim0.8$ for all fits. 

 The 2022 SED exhibits typical disk and Comptonization parameters, with the base model being reported as the best fit model with $AIC_c=7239.7$. We also find that the warm absorber component is negligible during this epoch (small column density) and that the addition of further absorption does not significantly improve the 2022 spectral fit. The addition of a partial coverer is a worse fit ($AIC_c=7249.1$).  The addition of a cold absorber does improve the fit statistic ($AIC_c=7217.6$), but produces highly unrealistic parameters requiring super-Eddington accretion ($\dot{m}/\dot{m}_{Edd}=1.55024$) and very large inner-disk radii ($R_{in}/\Rg=671.475 $). Therefore, additional absorption is not considered further in 2022.

The best-fit for the average 2022 spectrum (Figure \ref{fig:all_spec} with ratios in Figure \ref{fig:ratios}) requires an accretion rate of $\dot{m}/\dot{m}_{Edd}=0.013$, which is similar to the value of 0.03 obtained from the timing analysis by \citet{Gonzalez2024}. The measured black hole mass is $\log(M_{BH}/\Msun)=7.62$, which is slightly high, but within the range of the literature values \citep[e.g. ][]{Bentz2009, Bentz2010}. We note that better constrained parameters with errors are reported for the simultaneous multi-epoch fit in Section \ref{simfit}.

The most curious measurement is that of the accretion disk inner radius ($R_{in}$), for which we measure $R_{in}=55 \Rg$. This radius is at odds with the usual assumption that the inner radius of the accretion disk is at the innermost stable circular orbit ($ISCO$) \citep[e.g. ][]{Gonzalez2024}. Thus, we test a fit with the inner radius frozen to the ISCO, which finds $AIC_c=9748.8$, which is significantly poorer (especially in the UV band) than the best-fit with free $R_{in}$ ($\Delta AIC_c>2000$).  Large values for $R_{in}$ do persist in all the data.  Since it does provide a superior fit quality, we leave $R_{in}$ as a free parameter in this work and discuss the implications in Section \ref{disc}.

\subsection{ The average 2016 spectrum }\label{ave_2016}
The best-fit 2022 average model is used as the initial input for the average 2016 spectrum. For this SED, we fix certain parameters that we do not expect to vary between epochs to the best fit 2022 value. These include the Balmer continuum normalization, the black hole mass, and the host galaxy normalization. We first fit with no additional absorption and find a fit statistic of $AIC_c=2291.6$. 

This epoch is known to contain a partial covering eclipse \citep[][]{Gallo2021} along with the identification of several potential eclipses \citep{Pottie2023}.  Therefore, we consider the addition of a partial coverer. This model produces the most realistic parameters and the best-fit statistic \textcolor{black}{${AIC_c=2274.3}$}. Once again, the inner radius is measured to be a value much higher than is typically measured (\textcolor{black}{${R_{in}=106 \Rg}$}). We again test the assumption $R_{in}=R_{ISCO}$, which generates \textcolor{black}{${\Delta AIC_c=5642.4}$}. Leaving the inner radius free continues to be a significant fit improvement. The average SED along with the daily spectra (grey) are shown in the middle panel of Figure \ref{fig:all_spec}.

\subsection{ The average 2012 spectrum }\label{ave_2012}
We once again begin with the best-fit 2022 model and apply it to the 2012 average spectrum, fixing the same values as described for 2016. With no additional absorption we find $AIC_c=121.0$, while the addition of a cold absorber produces $AIC_c=118.5$. Replacing the cold absorber with the partial coverer produces $AIC_c=126.5$ and poorly constrained fit parameters. Therefore we report the inclusion of \textsc{ztbabs} to be the best model on the 2012 average spectrum. 

The 2012 value of the inner radius is again large. As with 2022 and 2016, we test fixing it at the  $ISCO$, which results in $AIC_c=6059.5$.  Therefore, leaving the inner radius free to vary is a significant fit improvement over fixing it to the ISCO. The bottom panel of Figure \ref{fig:all_spec} shows the average SED plotted over the 9 segments.

\subsection{The simultaneous fitting of 2012, 2016, and 2022 average spectra}\label{simfit}
In presenting a final model, the 2012, 2016, and 2022 average spectra are fitted together. \textcolor{black}{This allows us to better constrain parameters that do not vary significantly and are linked across the epochs}. The best fit average model at each epoch serves to guide the simultaneous fit.  Additionally, the BAT data are not considered in this fit.

The \textsc{xabs} ionization parameter, Balmer continuum normalization, Fe~K$\alpha$ line energy and width, black hole mass, host galaxy normalization, and the \textsc{xstar} ionization parameters and normalization are all linked between the epochs.  This fit finds parameter values that are largely comparable with the previous fits to individual epochs. 

\begin{figure}
    \centering
    \includegraphics[width=\columnwidth]{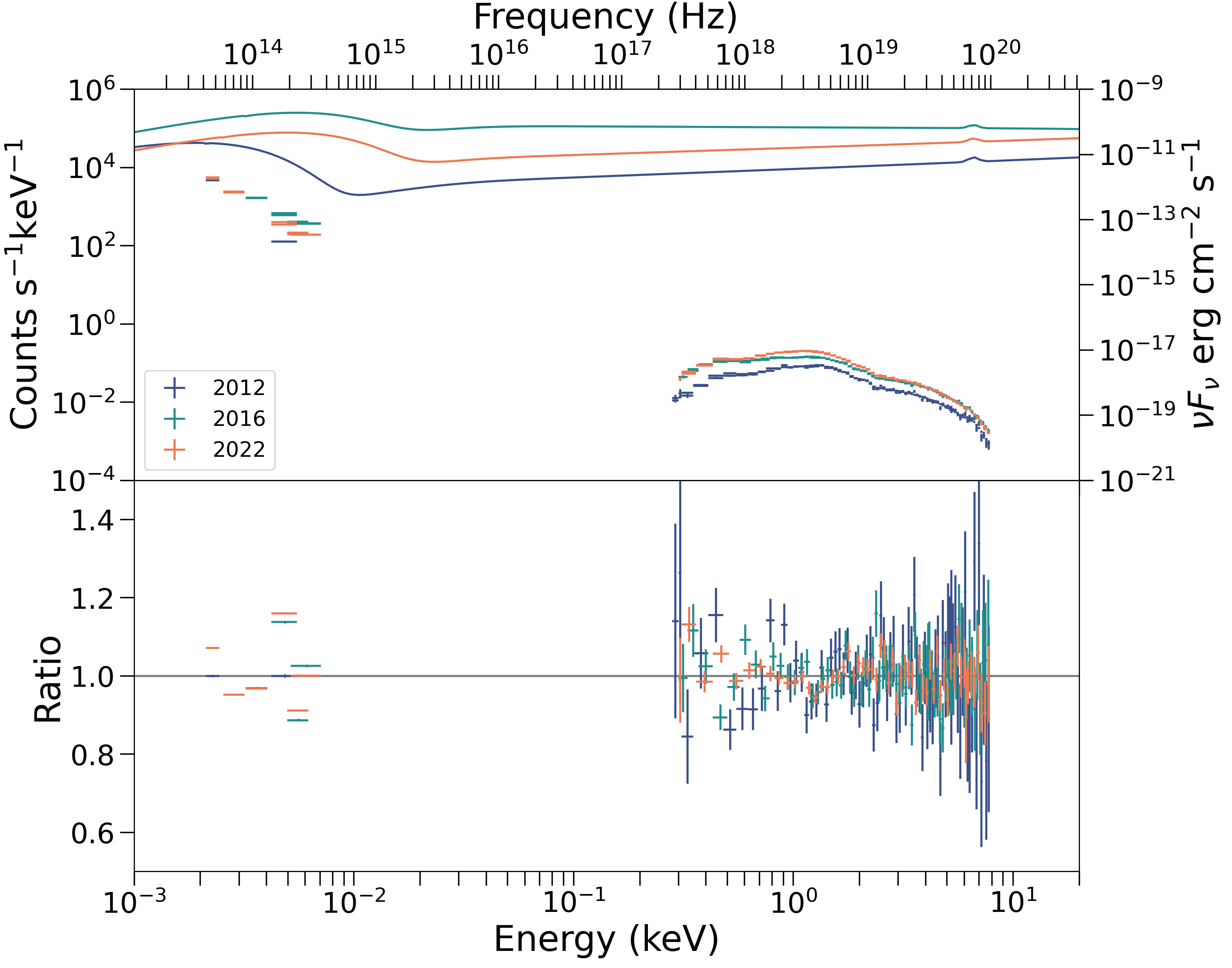}
    \caption{The complete average model applied to the three monitored epochs simultaneously. \textcolor{black}{In the upper plot, the data shown with error bars are plotted in count rate with the left axis. The intrinsic spectra in solid lines are plotted against the right axis, with the corresponding frequencies shown on the top axis.} The lower plot depicts the ratio of the data over the model.}
    \label{fig:sim_spec}
\end{figure}

To calculate errors on these parameters we use the \textsc{xspec} implementation of the Markov Chain Monte Carlo (MCMC) method.  This is performed using \textcolor{black}{100} walkers, a chain length of $100,000$, and a burn of $50,000$. \textcolor{black}{We note that the warm absorber column density is pegged to the maximum value ($10^{24}$\pscm), so we freeze it for the error calculations}. The full data and model are shown in Figure \ref{fig:sim_spec}.  The parameter values are reported in Table \ref{tab:model}.

\begin{table*}
	\centering
	\caption{The best fit model parameters found for the 2022, 2016, and 2012 average spectra, fit simultaneously. Tied parameters are indicated by a $t$.}
	\label{tab:model}
	\begin{tabular}{llccc} 
		\hline
        (1) & (2) &(3) & (4) & (5) \\
		Component & Parameter & 2022 Value & 2016 Value & 2012 Value\\
		\hline
        warm absorber (\textsc{xabs}) & $\log\xi/ \erg \cm^{-2} \s^{-1}$ & $\textcolor{black}{3.81\pm0.05}$ \   & $t$& $t$ \vspace{1mm} \\
         & $N_H/ 10^{24} \cm^{-2}$ & $
         \textcolor{black}{0.128^{+0.005}_{-0.049}}$ &$\textcolor{black}{1.0}$& $\textcolor{black}{0.02^{+0.03}_{-0.02}}$ \\  \hline
        Balmer continuum &  $norm$ & $\textcolor{black}{0.0130^{+0.0006}_{-0.0008}}$   &$t$&  $t$\\ \hline
         & $E/ \keV$ & $\textcolor{black}{6.62^{+0.07}_{-0.08}}$ & $t$  &$t$ \vspace{1mm} \\
        Gaussian emission (\textsc{zgauss}) & $\sigma/ \keV$ & $\textcolor{black}{0.3\pm0.2}$& $t$  & $t$ \vspace{1mm} \\
        & $norm/10^{-5}$ & $\textcolor{black}{7\pm2}$& $\textcolor{black}{1.2^{+0.2}_{-0.3}} $ &$\textcolor{black}{0.3\pm0.2}$ \\ \hline
         & $N_H/ 10^{24} \cm^{-2}$ & - & $\textcolor{black}{0.036^{+0.003}_{-0.004}}$ & - \vspace{1mm} \\
         partial coverer (\textcolor{black}{\textsc{xabs}})&$\log\xi/ \erg \cm^{-2} \s^{-1}$ & - &$\textcolor{black}{1.2^{+0.2}_{-0.3}}$ & - \vspace{1mm} \\
          & Covering Fraction & - & $\textcolor{black}{0.59\pm0.04}$& - \\\hline
          cold absorber (\textsc{ztbabs}) &$N_H/ 10^{22} \cm^{-2}$ & - & - & $\textcolor{black}{0.11\pm 0.02}$ \\ \hline
         coronal emission (\textsc{nthcomp})& $\Gamma$ & $\textcolor{black}{1.783^{+0.010}_{-0.007}}$ & $\textcolor{black}{2.03^{+0.08}_{-0.07}}$ &$\textcolor{black}{1.74^{+0.02}_{-0.03}}$ \vspace{1mm} \\

         & $norm/10^{-3}$ & $\textcolor{black}{9.82^{+0.09}_{-0.36}}$  & $\textcolor{black}{42^{+5}_{-4}}$  & $\textcolor{black}{2.21^{+0.09}_{-0.08}}$ \\ \hline
         &  $\log (M_{BH}/\Msun)$ &$\textcolor{black}{7.66\pm0.05}$ & $t$ & $t$ \vspace{1mm} \\
         disk emission (\textsc{ntdisk}) & $\dot{m}/\dot{m}_{Edd}$ & $\textcolor{black}{0.013\pm0.003}$ & $\textcolor{black}{0.09\pm0.02}$  & $\textcolor{black}{0.03\pm0.01}$\vspace{1mm} \\
         & $R_{in}/\Rg$ &  $\textcolor{black}{58\pm7}$ &  $\textcolor{black}{100\pm10}$ &  $\textcolor{black}{270\pm30}$ \\
        \hline
        photoionized emitter (\textsc{xstar}) & $\log\xi/\erg \cm^{-2}\s^{-1}$ & $\textcolor{black}{1.5\pm0.2}$  & $t$ &$t$ \vspace{1mm}\\
         & $norm/10^{-5}$ & $\textcolor{black}{4.0^{+0.5}_{-0.4}}$ & $t$ & $t$\\ \hline
         host galaxy & $norm/10^{-15}$ & $\textcolor{black}{9.3\pm0.1}$ & $t$ &$t$ \\ \hline
	\end{tabular}
\end{table*}

\section{Fitting the daily variations in the optical-to-X-ray SEDs }\label{all_daily}

In this section, we examine the approximately daily variations in the SED that are taking place over the $\sim 60-76$~day monitoring campaigns.

\subsection{Fitting the daily spectra in 2022}\label{daily_2022}
We apply our best fit simultaneous model to the individual days of the 2022 observations. We first freeze several parameters which we do not expect to vary on short time scales or that the data are insensitive to. \textcolor{black}{These include all parameters relating to the warm absorber, the Balmer continuum, the Gaussian line, the photoionized emitter, and the host galaxy.} We also freeze the black hole mass to the average value. 

After fitting the first day, the fitting is performed sequentially so that the input model for each day is the best-fit model from the previous day.  This model and procedure yields reasonable fits for all spectral. The mean parameters for the daily models are reported in column 3 of Table \ref{tab:model_daily}. The mean daily parameters are found to be consistent with the fit to the average spectrum in the previous section. All observed days in 2022 utilize the 6 UVOT filters with the exception of one (MJD 59822) for which we are missing UVW2 and UVM2. To account for this, the disk parameters are fixed to the values of the previous day (MJD 59821).

\subsection{Fitting the daily spectra in 2016}\label{daily_2016}
The 2016 daily fitting is performed in the same way as described for 2022, we begin from the simultaneous model with a few additional frozen parameters and fit the days sequentially. We note that there are several days during which not all UVOT filters are used (U, UVW1, UVW2, and UVM2 for 2016). For these days, if the model parameters are significantly different to nearby days, we fix the disk parameters to be those of one of the adjacent days. In addition, there are several days during which the data are insufficient to model, thus we combine data from these days with adjacent ones. We report the mean values for the daily fits in column 4 of Table \ref{tab:model_daily}. Once again the mean values of the daily models are in good agreement with the average best fit model.

\subsection{Fitting segmented spectra in 2012}\label{segments_2012}
The 2012 segmented spectra are fit simultaneously, starting from the average model in Section \ref{ave_2012} that includes the cold absorber. Many parameters are fixed as in Section~\ref{daily_2022}, leaving \textsc{ztbabs} $N_H$, $\Gamma$, $norm_{nthcomp}$, $\dot{m}$, and $R_{in}$ free to vary between segments. The overall fit quality for the 9 combined segments is $AIC_c=728.5$ . This is the best statistical fit for the 2012 segments, however, the disk values are inconsistent with the expected values and the observed lower flux during this epoch. 

Next, we linked the accretion rate between all segments. This resulted in an accretion rate value that was comparable with the average model ($\dot{m}/\dot{m}_{Edd}=0.02988$), and $AIC_c=1037.1$. We note that the only 2 UVOT filters available during the 2012 observations are V and UVW1. The V band is dominated by host galaxy emission, thus we effectively have only one filter to constrain the disk parameters, for this reason we favour the model with the tied accretion rate. The mean parameter values from the 2012 segmented fits are reported in column 5 of Table \ref{tab:model_daily}, all parameters are in agreement with the average model.

\begin{table*}
	\centering
	\caption{The mean parameters computed for the daily spectra in each of the monitored epochs. Dashes indicate that a component is not used for that epoch, parameters with a superscript $t$ are tied between all individual spectra. The reported errors for 2022 and 2016 are propagated from the fit errors assuming Gaussian statistics and the 2012 errors are calculated as the standard deviation of the parameter.}
    	\label{tab:model_daily}
	\begin{tabular}{ccccc} 
		\hline
        (1) & (2) &(3) & (4) & (5) \\
		Component & Parameter & Mean 2022 Daily Value & Mean 2016 Daily Value & Mean 2012 Segment Value\\
		\hline
         
          partial coverer (\textcolor{black}{\textsc{xabs}})& Covering Fraction & - & \textcolor{black}{$0.51\pm0.06$} & - \\\hline
          cold absorber (\textsc{ztbabs}) &$N_H/ 10^{22} \cm^{-2}$ & - & - &$ 0.2\pm0.1$\\ \hline
         coronal emission(\textsc{nthcomp})& $\Gamma$ & $1.78\pm0.01$ & \textcolor{black}{$2.00\pm0.05$}  &$1.75\pm0.07$  \\
         & $norm/10^{-3}$ & $11.05\pm0.09$  & \textcolor{black}{$51\pm4$}  &  $2\pm1$ \\ \hline
         & $\dot{m}/\dot{m}_{Edd}$ & $0.0139\pm0.0001$ &\textcolor{black}{$0.090\pm0.003$  }& $0.02988^t$\\ 
         disk emission (\textsc{ntdisk})& $R_{in}/\Rg$ &  $58\pm1$ &  \textcolor{black}{$109\pm6$} &  $282\pm56$    \\
        \hline
	\end{tabular}
\end{table*}

\section{Correlation analysis}\label{cor}

We examine correlations between the various model parameters resulting from the yearly and daily spectral fit analysis in Section \ref{model} and \ref{all_daily}, respectively.   This also includes \textcolor{black}{model independent} parameters like the luminosities and $\Delta \alpha_{ox}$ values as described in Section \ref{lc}.  

The correlations are quantified using the Pearson correlation coefficient. We present these correlation coefficients as well as the scatter plots with linear best-fits and histograms for each pair of parameters for all three monitored epochs in Figure \ref{fig:cor_plot}.  Recall, the accretion rate ($\dot{m}$) is linked across all the segments in 2012, thus for completeness the correlation coefficients calculated without including the 2012 accretion rate are reported in brackets in Figure \ref{fig:cor_plot}. This plot is inspired by the \textsc{panels.pairs} function from the \textsc{psych} package in R \citep[][]{Revelle2024}.

\subsection{Correlations in Yearly Parameters}\label{year_cor}

\begin{figure*}
    \centering
    \includegraphics[width=\textwidth]{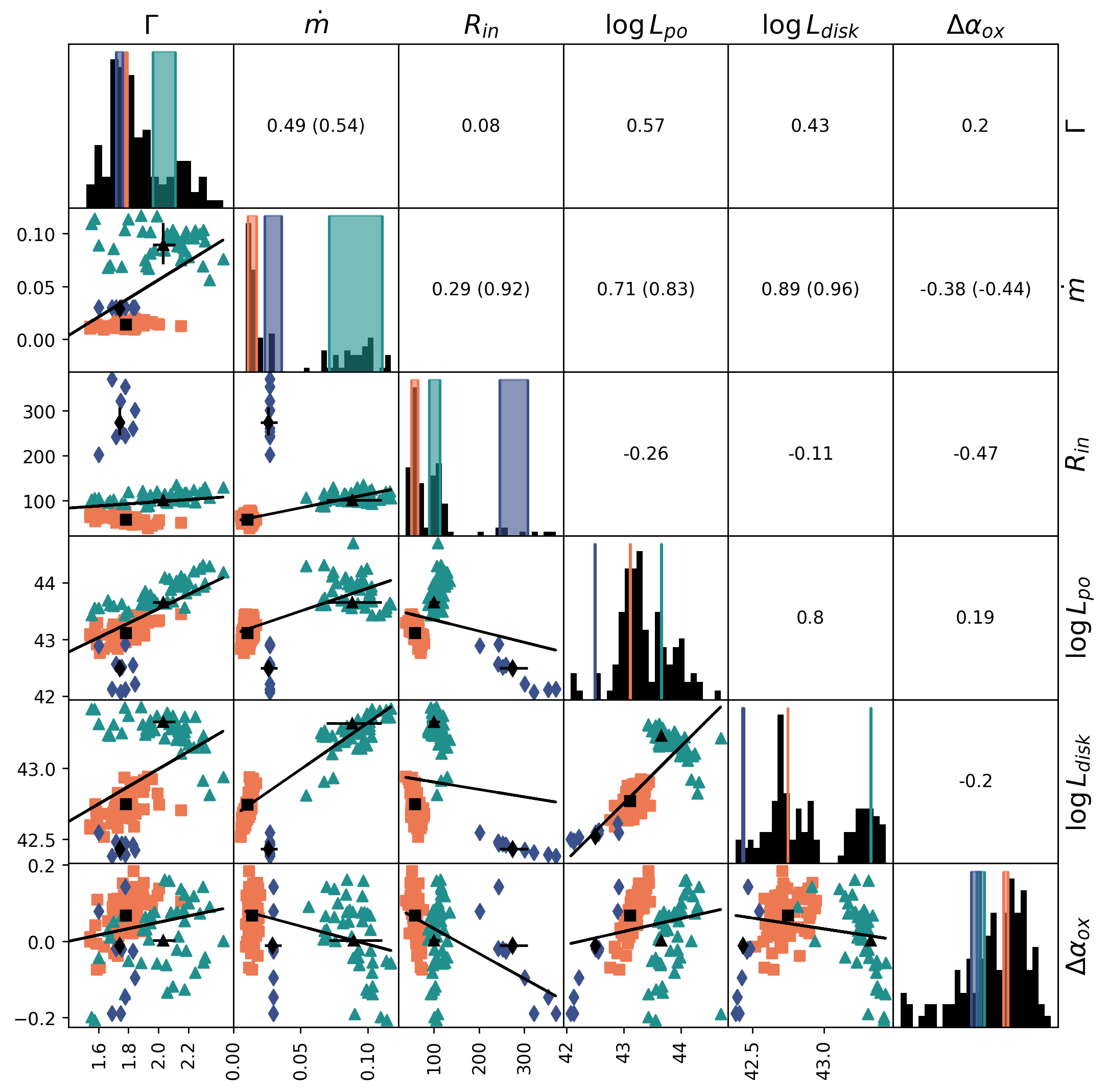}
    \caption{Lower left: Scatter plots for all parameters shared between the 2012 segmented spectra and 2016 and 2022 daily spectra, indicated by blue diamonds, teal triangles, and orange squares respectively. The solid black lines indicate the linear best fit to the data. Solid black shapes indicate  the values measured by the simultaneous average fit. Upper right: Pearson correlation coefficients for each pair of parameters, values in brackets are calculated while neglecting the 2012 tied accretion rate. Diagonal: Histograms of each parameter, coloured bars represent values measured by the simultaneous average fit, with the width determined by the errors.}
    \label{fig:cor_plot}
\end{figure*}

The long-term correlations measured by simultaneously fitting the average spectrum in 2012, 2016 and 2022 (Section \ref{model}) are shown as black filled symbols in Figure \ref{fig:cor_plot}.  While only three epochs are compared, much of the behaviour appears as expected.

The power law photon index and luminosity exhibit the well-known steeper-when-brighter trend \citep[e.g.][]{Markowitz2003, Ursini2016}; the disk luminosity and Eddington accretion rate are positively related \citep[][]{Peterson1997}; and the luminosity of the power law and disk are well-correlated \citep[e.g.][]{Lusso2016}.  The remaining pairs of parameters appear to have more scatter.  More long-term monitoring would be necessary to provide significant insight on these long time scales.

\subsection{Correlations in Daily Parameters}

Examining the relations in the daily variations reveals more complex behaviour.  For example, there are instances when the correlations are strong at one epoch and not another, or occasions where the relationships change.  The relationships are plotted for all epochs in Figure \ref{fig:cor_plot} and for each individual epoch in Appendix \ref{cors_app}.

As with the long-term correlations (Section \ref{year_cor}), with the short-term variability the correlations between disk luminosity and accretion rate are evident in 2022 and 2016 (not measurable in 2012).  Similarly, the steeper-when-brighter relation is evident in 2022 and 2016, though inconclusive in 2012 with fewer data points.  Curiously, the positive correlation between disk and corona (power law) luminosity exists in 2022 ($r=0.53$) and 2012 ($r=0.89$), but in 2016, there is a strong anti-correlation between the two parameters ($r=-0.71$).

From the spectral fitting, the best-fit inner disk radius ($R_{in}$) was measured to be high at all three epochs ($R_{in} \approx 60, 100, \rm and~ 275\Rg$ in 2022, 2016, and 2012, respectively), perhaps indicative of a non-standard disk.  Of further interest is the variability behaviour of this parameter.

In 2022 and 2012, $R_{in}$ is strongly anti-correlated with the disk luminosity, accretion rate, corona luminosity, and the X-ray weakness parameter.  Additionally, in 2012, there is a strong correlation between $R_{in}$ and the level of cold absorption ($n_H$).  Consequently, in 2012, there is a strong anti-correlation between $n_H$ and the disk luminosity, accretion rate, corona luminosity, and the X-ray weakness parameter.  There was no cold absorption present in 2022.

In contrast, in 2016, the correlations with $R_{in}$ were less significant. However, it is notable that the anti-correlation with disk luminosity persists ($r=-0.43$) as in 2012 and 2022, but the correlation with corona luminosity is positive ($r=0.49$).  These differences might suggest that \ngc{} is exhibiting different physical behaviour in each of the three epochs.

\section{Discussion}\label{disc}

\subsection{Important time scales and long-term variability}

The variations in the optical-to-X-ray SED of \ngc{} are examined on yearly time scales between 2012 and 2022, as well as on nearly daily time scales for durations of $\sim60-75$~days in 2012, 2016, and 2022.  Assuming the temperature profile for a standard accretion disk and the values in Table ~\ref{tab:model_daily}, the black body emission in the UVW1 filter (2600 \AA) originates from approximately $93\Rg$ \citep[following e.g.][]{Gallo2018}.  Further, assuming a viscosity parameter of $\alpha_\nu=0.1$ and vertical-to-radial disk ratio of $h/r=0.01$, at this distance, we estimate the dynamical time-scale $t_{dyn}\approx15$ 
days, a thermal time-scale $t_{th} \approx 150$ days, and a viscous time-scale $t_{vis} \approx 1.5 \times10^6$ days.  The light-travel time across this region is $t_{lc}\approx0.2$ days.  The year-to-year variations we examine in this work allow us to probe thermal time scales in the accretion disk, whereas the more rapid variability examined during the daily monitoring campaigns reflect changes on dynamical time scales.

The year-to-year changes in \ngc{} appear consistent with a standard accretion disk and typical AGN behaviour.  The Eddington accretion rate fluctuates between $\dot{m}/\dot{m}_{Edd} \approx 0.01 - 0.09$ on thermal time scales.  The UV-to-X-ray spectral slope, $\alpha_{ox}$, follows the trends of being more X-ray weak with increasing UV luminosity \citep[e.g.][]{Strateva2005,Just2007}.  Finally, the power law emission associated with the corona exhibits the well-known steeper-when-bright relation.

\subsection{The large inner-disk radius}

Pointed X-ray observations of \ngc{} show that the AGN displays rapid variability, typical of type I Seyferts, and consistent with a compact corona \citep[e.g.][]{Tennant1981, Fiore1992, Gallo2021,Gonzalez2024}.  Estimates of the corona size during the eclipse suggest that the region is compact and on the order of $\sim20\Rg$ \citep[][]{Gallo2021}.

On the other hand, throughout the SED fitting, we measure values of the accretion disk inner radius that are uncharacteristically high between $\sim60\Rg$ (in 2022) and $\sim275\Rg$ (in 2012).  These measurements are robust and independent of the continuum model used. In addition to our chosen \textsc{ntdisk+nthcomp}  model,  the tested models include the self-consistent \textsc{agnsed} \citep[][]{Kubota2018}; the combination of \textsc{diskbb} and \textsc{nthcomp}; the combination of a broken power law and \textsc{ntdisk}; and the combination of \textsc{simpl} \citep[][]{Steiner2009} with \textsc{ntdisk}. \textcolor{black}{We note that the \textsc{ntdisk+nthcomp} model was preferred because of the examined parameters and the fast computing time.  The results and statistics were comparable with the different model combinations. }

\textcolor{black}{A truncated disk is expected in low-luminosity AGN (LLAGN) where the inner part of a standard disk is replaced by a radiatively inefficient flow \citep[e.g.][]{Narayan1994,Lasota1996}.  However, \ngc{} does not appear like a LLAGN \citep[e.g.][]{Ptak2004,Younes2019}.  The high-amplitude variability and modest Eddington accretion rate suggest it is more likely a normal Seyfert galaxy.  A truncated inner disk at $\sim10\Rg$ was reported for the narrow-line Seyfert 1 galaxy WKK 4438 by \citet{Gallo2022}.  That conclusion was based on modelling of the blurred reflection component rather than the SED, but points to a similar scenario in another type I AGN.}

The large radius is unusual for a standard accretion disk, and may not agree with the $R^{-3/4}$ temperature profile we anticipate, however, the observations may not be inconsistent with the existing timing analyses of NGC~6814 \citep[][]{Troyer2016,Gonzalez2024}.  Based on the 2022 light curves, \citet{Gonzalez2024} measured a time delay between the X-ray and UVW2 (2078 \AA{}) of \textcolor{black}{$0.39^{+0.13}_{-0.08}$} days, which corresponds to the light travel time across $\sim153\Rg$. \textcolor{black}{This is consistent with the time delay between the X-ray and UVW1 (2600 \AA{}) found by \citet{Troyer2016} for the 2012 epoch.}  A large inner disk radius appears consistent with the absence of any strong blurred reflection \citep[e.g.][]{Gallo2021} or X-ray reverberation signals (Hodd et al. in prep).

If we accept that the inner disk edge can be extended, then of interest is what has caused it to move $\sim200 \Rg$ in four years.  This could occur on viscous time scales as the accretion flow changes.  There are indications of this occurring in the SED models.  However, the estimated viscous time scales are very long (i.e. $\sim10^6$ days) unless we consider a disk thickness that is $\sim30$-times greater than what we have assumed for the standard disk (i.e. $h/r\sim 0.3$).  An ``inflated’’ disk surrounding a compact corona might be consistent with the obscuration events / eclipses \citep[e.g.][]{Leighly1994, Gallo2021} and changing-look nature \citep[][]{Sekiguchi1990,Jaffarian2020} of \ngc{}. 

\subsection{The eclipsing epoch}

The X-ray eclipse of \ngc{} captured in its entirety by \xmm{} \citep[][]{Gallo2021} occurred during the 2016 \swift{} monitoring campaign.  \citet{Pottie2023} predicted that several eclipses like the one captured by \xmm{} might have occurred during that monitoring campaign.

In comparing the SEDs and behaviour at each epoch, we note that 2012 and 2022 are comparable.  There is a difference in the measured accretion rate, inner disk radius, and level of cold absorption; however, the variability behaviour of the parameters is reasonably similar.  The 2016 campaign does show marked differences compared to 2012 and 2022, as noted in Section \ref{pca}.  There are also differences in the relations between parameters (Section \ref{cor}) and 2016 shows the strongest evidence for a warm absorber.  We can only speculate if any of these differences in 2016 generate eclipses or are consequences of eclipses.

\section{Conclusion}\label{conclusion}

We have examined the changes in the optical-to-X-ray SED of NGC~6814 using Swift data at five epochs within a 10-year period, including three epochs of high-cadence monitoring in 2012, 2016, and 2022.  The 2016 monitoring captures the \xmm{} eclipse which has been thoroughly examined.

\begin{itemize}
\item The X-ray PCA points to continuum variations in the photon index and normalization of the power law driving the 2012, 2016 and 2022 variability, with the eclipse having a significant effect on the shape of the first principal component in 2016.  The low-flux levels in 2016 and 2012 show similar X-ray weakness, but the PCA results confirm that the origin of these low-flux levels are not the same.
    \item SED fitting reveals differing absorption across the three epochs.  The ``typical’’ state observed in 2022 has variability consistent with continuum changes, and has no significant absorption.  The 2016 eclipsing state is consistent with some continuum variability with the addition of a partially covering absorber.  We find the 2012 low-flux state to be consistent with variability  generated mostly through continuum changes, with small contributions from a cold absorber.
    \item The multi-epoch SED models are consistent with a black hole ($M_{BH}\approx 10^{7.6} \Msun{}$) that is accreting between $\dot{m}\approx0.01-0.1$.  While the corona (primary X-ray source) is compact, all epochs are better fit with an accretion disk inner radius that is much larger than the ISCO, implying the possibility of a non-standard accretion disk or central structure in \ngc{}.
    
\end{itemize}

\section*{Acknowledgements}
The authors thank the anonymous referee for providing a thoughtful review. This work made use of data supplied by the UK Swift Science Data Centre at the University of Leicester.  LCG acknowledges financial support from the Natural Sciences and Engineering Research Council of Canada (NSERC) and the Canadian Space Agency (CSA).

\section*{Data Availability}
The data presented here is publicly available through the NASA HEASARC Archive (\url{https://heasarc.gsfc.nasa.gov/docs/archive.html}) and Swift observatory websites (\url{https://www.swift.ac.uk/index.php}).


\bibliographystyle{mnras}
\bibliography{ref}



\appendix
\section{Individual epoch correlations}\label{cors_app}

\begin{figure*}
    \centering
\includegraphics[width=\textwidth]{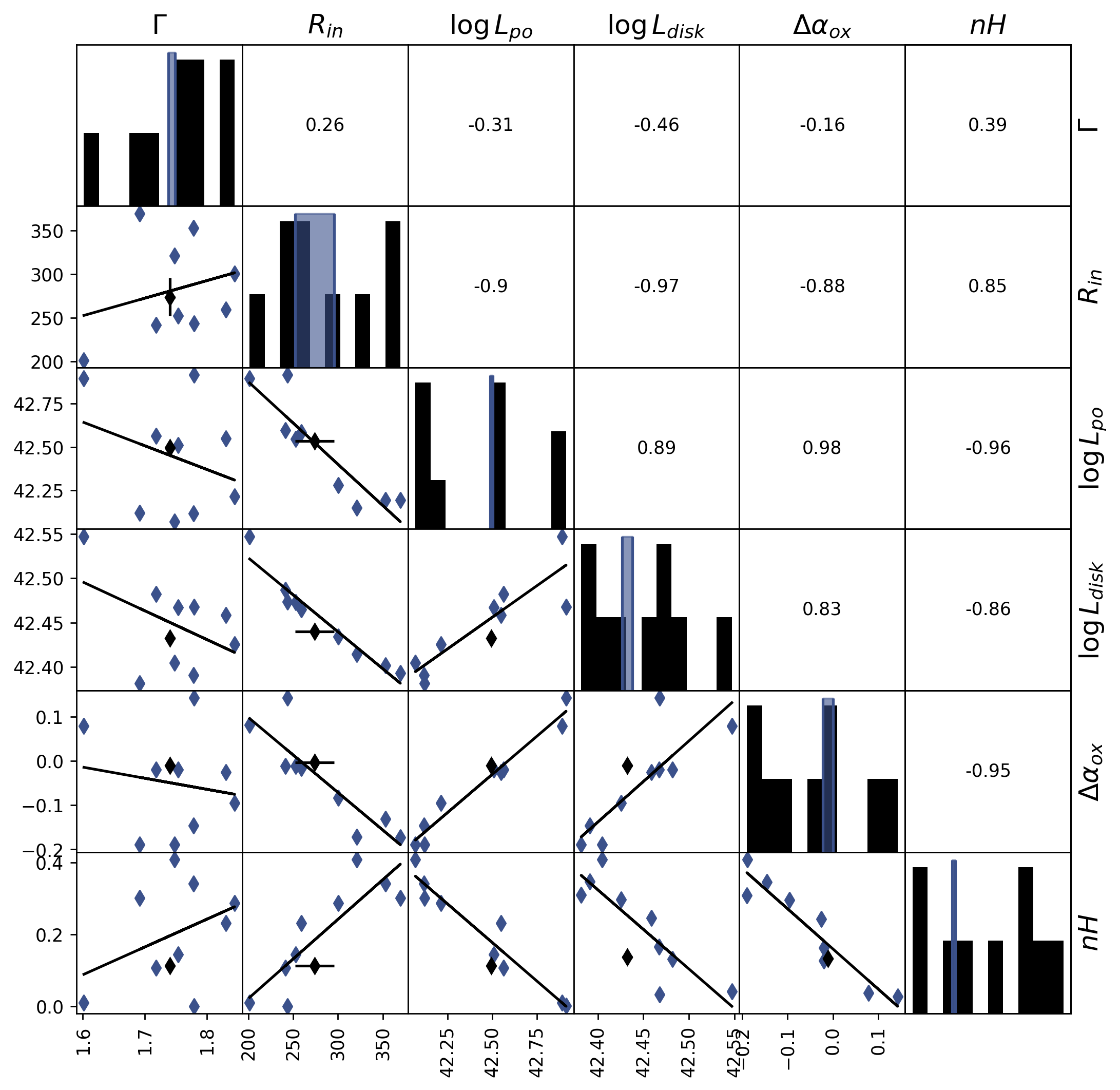}
    \caption{Lower left: Scatter plots for all 2012 parameters. The solid black lines indicate the linear best fit to the data. Solid black shapes indicate  the values measured by the simultaneous average fit. Upper right: Pearson correlation coefficients for each pair of parameters. Diagonal: Histograms of each parameter, coloured bars represent values measured by the simultaneous average fit, with the width determined by the errors.}
    \label{fig:cor_2012}
\end{figure*}

\begin{figure*}
    \centering
\includegraphics[width=\textwidth]{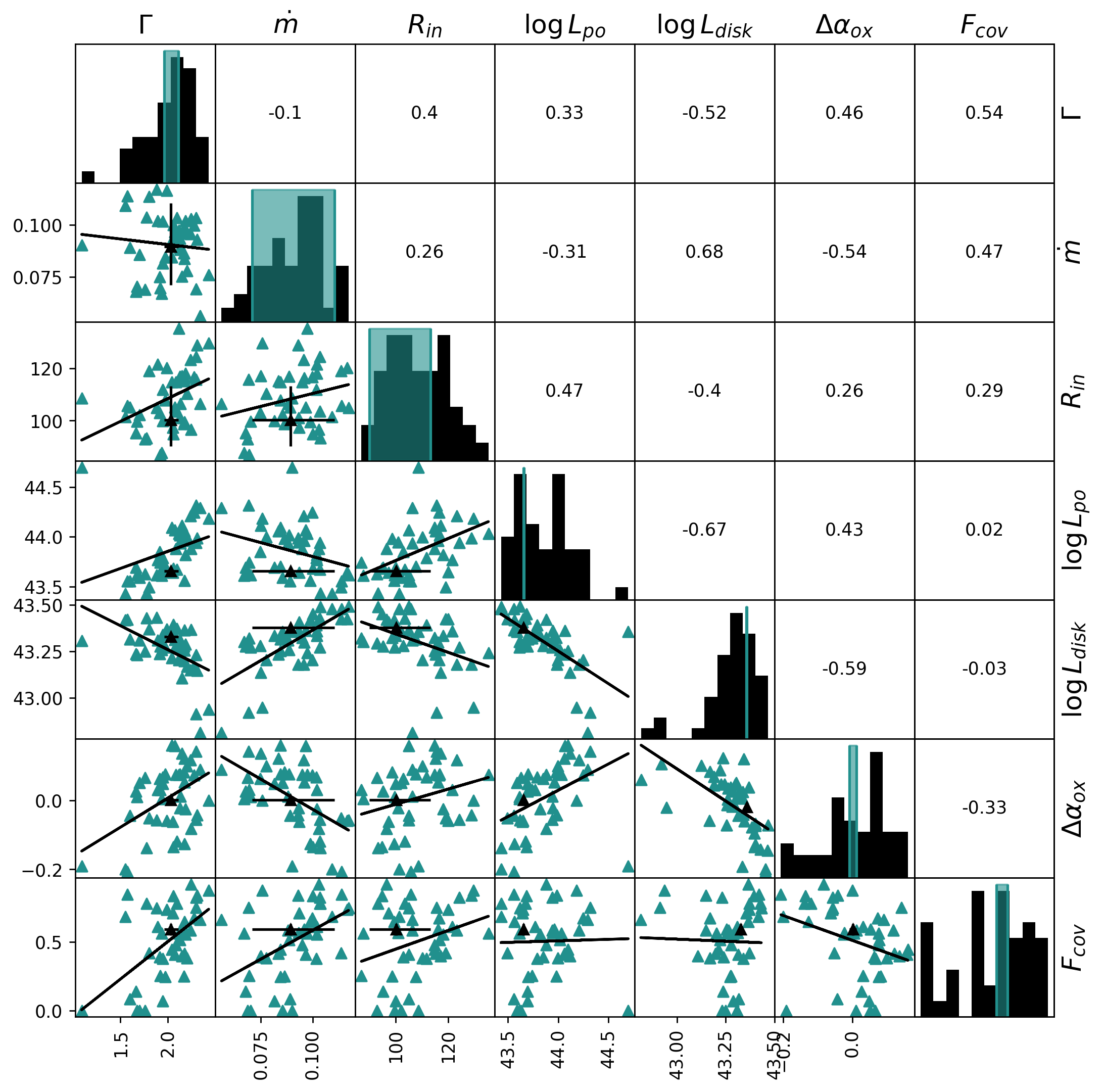}
    \caption{Lower left: Scatter plots for all 2016 parameters. The solid black lines indicate the linear best fit to the data. Solid black shapes indicate  the values measured by the simultaneous average fit. Upper right: Pearson correlation coefficients for each pair of parameters. Diagonal: Histograms of each parameter, coloured bars represent values measured by the simultaneous average fit, with the width determined by the errors.}
    \label{fig:cor_2016}
\end{figure*}

\begin{figure*}
    \centering
\includegraphics[width=\textwidth]{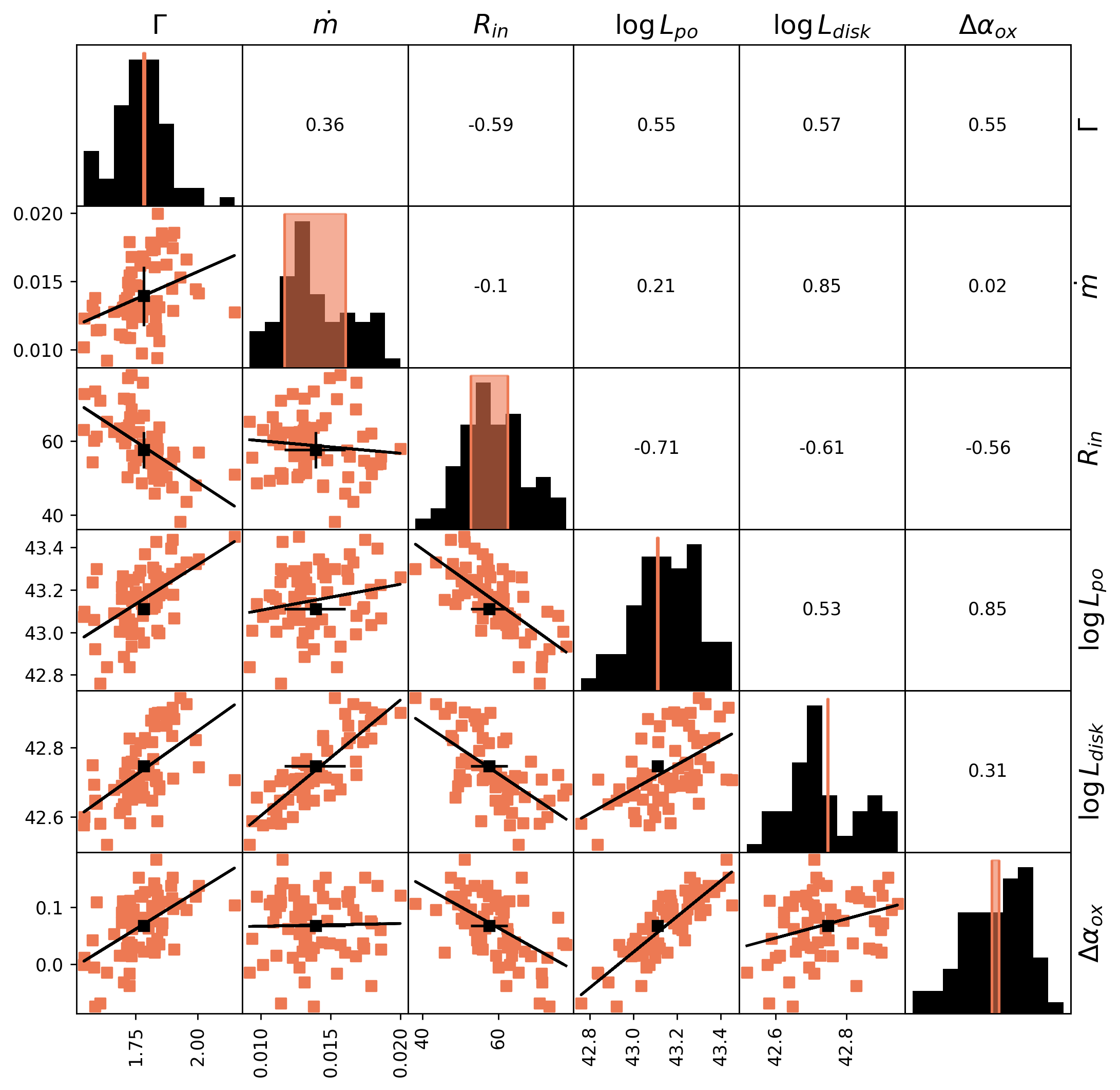}
    \caption{Lower left: Scatter plots for all 2022 parameters. The solid black lines indicate the linear best fit to the data. Solid black shapes indicate  the values measured by the simultaneous average fit. Upper right: Pearson correlation coefficients for each pair of parameters. Diagonal: Histograms of each parameter, coloured bars represent values measured by the simultaneous average fit, with the width determined by the errors.}
    \label{fig:cor_2022}
\end{figure*}


\bsp	
\label{lastpage}
\end{document}